\documentclass[utf8]{FrontiersinHarvard} 

\usepackage{url,hyperref,microtype,subcaption}
\usepackage[onehalfspacing]{setspace}

\def\keyFont{\fontsize{8}{11}\helveticabold }
\def\firstAuthorLast{Vaida {et~al.}} 
\def\Authors{David D Vaida\,$^{1,*}$ and Ryan Jeffrey Farber\,$^{2}$}
\def\Address{$^{1}$ Independent Researcher \\
$^{2}$Department of Physics, Purdue University Fort Wayne, Fort Wayne, IN, USA}
\def\corrAuthor{David D Vaida}

\def\corrEmail{ddvaida28@gmail.com}

\begin{document}
\onecolumn
\firstpage{1}

\title[Little Red Dots in the JWST Era]{
Little Red Dots: The Assembly of Early Supermassive Black Holes in the JWST Era
} 

\author[\firstAuthorLast ]{\Authors} 
\address{} 
\correspondance{} 

\extraAuth{}

\maketitle

\begin{abstract} 
Since the launch of James Webb Space Telescope (JWST) in late 2021, our understanding of high-redshift objects has faced several upheavals. JWST has discovered much more massive galaxies and supermassive black holes (SMBH) than cosmological models had expected. Furthermore, JWST observations have revealed an entirely novel population of high-redshift objects. Characterized by a dominant red rest-frame component and point-like morphology, these ``little red dots'' (LRD) have set off a flurry of observational and theoretical follow-up. The current identity of LRD is highly debated, yet falling into two main scenarios: active galactic nuclei (i.e., SMBH) or compact star-forming regions. If star-forming, LRD would represent the highest stellar densities ever observed. If SMBH, their high Eddington fractions, and already high masses, help elucidate the growth of the most massive SMBH found by JWST in the early Universe ($z \gtrsim4)$. In this mini-review, we present the observational evidence accumulated to date, including sub-millimeter probes of LRD dust masses, constraints on radio and X-ray emission from stacking, and rest-frame ultraviolet \& optical measurements provided by JWST. Furthermore, we highlight how identifying additional LRD that are truly primarily SMBH-driven may help to shed light on the formation of `overly massive' SMBH discovered by JWST within the first billion years since the Big Bang.

\tiny
 \keyFont{ \section{Keywords:} Little Red Dots, Supermassive Black Holes, JWST, AGN, Super-Eddington Accretion, Supermassive Stars, Direct Collapse Black Holes, Primordial Black Holes} 
\end{abstract}

\section{Introduction}
Since the launch of James Webb Space Telescope (JWST; \citealt{gardner2023james,mcelwain2023james}) $\sim$four years ago, JWST has discovered an increasing population of supermassive black holes (SMBH) that challenge the current paradigm of black hole formation and growth (\citealt{jacak2025possible} and see the recent review \citealt{harikane2025early}). 
Specifically, JWST-discovered SMBH are far more massive than expected to be present at such high redshifts ($\sim$billion solar mass black holes are present within the first billion years after the Big Bang), suggesting much more rapid formation than would be possible for Pop III stars accreting at the Eddington limit (\citealt{kiyuna2025super} and see the $\sim$recent reviews \citealt{inayoshi2020assembly,volonteri2021origins,jeon2025physical}). 

Instead, these `overly' massive SMBH may have formed through Eddington-limited accretion from heavy seeds, 
such as
direct collapse black holes from atomic cooling halos and supermassive stars \citep{kiyuna2024sequential,lu2024direct,jeon2025little}, 
or primordial black holes \citep{delos2024structure,riotto2025future,carr2025history}. 
Alternatively, these SMBH may have formed through hierarchical merging of light seed black holes in active galactic nucleus (AGN) disks \citep{vaccaro2024impact}, in dense nuclear or globular star clusters \citep{kritos2025supermassive,lahen2025mergers} or had their growth rates boosted beyond the Bondi rate (and Eddington limit) via fuzzy dark matter soliton cores \citep{chiu2025boosting}. 

Clearly, much uncertainty exists regarding precisely how SMBH grew so rapidly in the early universe. Perhaps the best way to uncover the physical mechanism by which these black holes formed is to find the extension of such a population at high-redshifts that are growing at lower luminosities. Such an intermediate mass population will help to illuminate the seed masses \citep{akins2025cosmos}; moreover, constraints on the accretion rates of such intermediate objects will help understand what seed masses are required for the requisite rapid growth to the observed SMBH masses.

Possibly constituting such intermediate mass SMBH in an active state, little red dots (LRD) describe a typically high-redshift population with compact morphology, broad $\sim$1000s km/s H$\alpha$ line widths \citep{kocevski2024rise, zhang2025analysis}, and simultaneously blue ultraviolet (UV) continua yet red optical colors \citep{setton2024little, hviding2025rubies}. Originally, LRD were detected for redshifts $4.2 < z < 5.5$ \citep{matthee2024little}; however, additional JWST observations have uncovered a larger population of LRD extending from redshifts $3 < z < 10$ \citep{labbe2024uncover,graham2025dot}. In addition to a growing population of LRD characterized by JWST, LRD have been investigated by increasingly deep multiwavelength campaigns stretching from radio to X-ray bands.


In this mini-review, we highlight the findings of the most recent LRD observations across the electromagnetic spectrum with a focus on identifying the physical origins of LRD. In section \ref{sec:obs} we contextualize the discovery of LRD with respect to the first few JWST surveys performed. Subsequently, in section \ref{sec:obsConstraining}, we discuss additional multi-wavelength studies that round out the characterization of LRD. 
Synthesizing the observations detailed in prior sections, in section \ref{sec:interp}, we discuss theoretical interpretations of LRD that claim to explain the observed properties of LRD. Finally, we zoom-out and discuss the state of the field of LRD origins more broadly, along with the future directions that can more conclusively end the controversy in section \ref{sec:disc_and_conclude}. Understanding these mechanisms is crucial for accurately interpreting high-redshift observations and for uncovering how the first SMBHs originated.

\section{LRD Discovery through the Lens of JWST Surveys}
\label{sec:obs}
Some of the first images from JWST, utilizing the Near Infrared Camera \citep[NIRCam;][]{rieke2023performance}, through the Cosmic Evolution Early Release Science program \citep[CEERS;][]{finkelstein2023ceers}, were found to contain very red galaxies (e.g., \citealt{endsley2023jwst,labbe2023population,onoue2023candidate}) implying high-redshift galaxies much more massive than expected from the current cosmological paradigm $\Lambda$ Cold Dark Matter (CDM; \citealt{boylan2023stress,ferrara2023stunning,lovell2023extreme}. 

Consideration of the longer wavelength properties of these red objects with the Mid-Infrared Instrument \citep[MIRI;][]{argyriou2023jwst} and spectra collected with the Near-Infrared Spectrograph \citep[NIRSpec;][]{bagnasco2007overview} resulted in lower mass estimates, while identifying very broad Balmer line emission and a V-shaped spectral energy distribution (SED): simultaneous blue rest-frame UV and red rest-frame optical colors \citep{carnall2023massive,kocevski2023hidden,barro2024extremely}. 
However, extremely large equivalent widths of the Balmer emission lines suggested a stellar component may contribute a significant fraction to the otherwise clear signature of AGN activity \citep{carnall2023massive}.

Although a small fraction of these red objects were later found to be foreground brown dwarfs, a larger sample of low-luminosity red objects at high-redshift with a characteristic v-shaped SED, point-like morphology, and broad Balmer lines \citep{furtak2023jwst,greene2024uncover,labbe2024uncover} were discovered through the Ultradeep NIRSpec and NIRCam Observations before the Epoch of Reionization \citep[UNCOVER;][]{bezanson2024jwst}) program. 
Combining deep NIRCam photometric and wide-field slitless spectroscopy (WFSS; grism) observations from the EIGER \citep{kashino2023eiger} and FRESCO \citep{oesch2023jwst} surveys, \citet{matthee2024little} coined the term `little red dots' (LRD) to explain the surprisingly abundant population: 
about 1\% of galaxies at redshift $z > 5$ were found to host LRD, while the number density $10^{-5}$ per cMpc$^{-3}$ (co-moving cubic Mega-parsec) is 10-100 times larger than expected by extrapolating the UV luminosity function.


Following up on the primarily photometric surveys that incidentally contained little red dots, the Red Unknowns: Bright Infrared Extragalactic Survey (RUBIES; \citealt{de2025rubies}), was designed to specifically investigate LRD through a homogeneous spectroscopic survey. Utilizing NIRSpec and covering cosmic noon ($z \sim$2) and earlier, RUBIES has revealed that LRD show a spectroscopic link between three defining characteristics: broad Balmer lines, a rest-optical point source, and a v-shaped continuum, across 1500 galaxies studied \citep{hviding2025rubies}.

\begin{figure}[ht]
\begin{center}
\includegraphics[width=\textwidth]{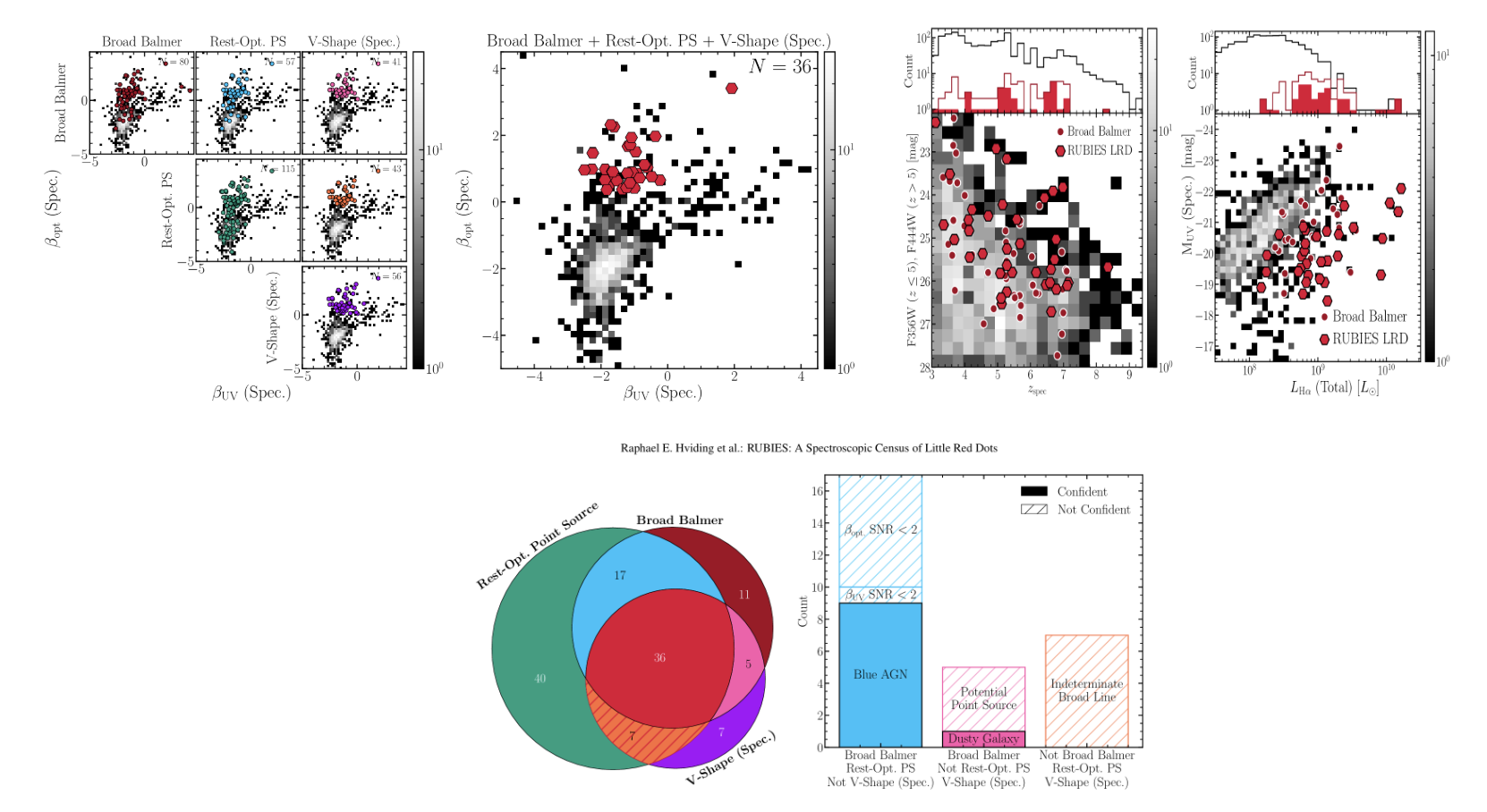}
\end{center}
\caption{
Adapted from \citet{hviding2025rubies}, their Figures 6, 7, and 8 to composite the RUBIES diagnostics for Little Red Dots. 
In the upper left `corner' plot and upper-middle plot, the $\beta_{\text{UV}}$--$\beta_{\text{opt}}$ maps show where sources with broad Balmer lines, unresolved rest-optical point sources, and V-shaped continua sit relative to all $z_{\text{spec}} > 3.1$ objects. 
In the upper right, the two heatmaps respectively show the photometric flux in two filters for high vs. low redshift sources (threshold $z = 5$) as a function of spectroscopic redshift, and the spectroscopically derived UV absolute magnitude versus the H$\alpha$ luminosity. Those two heatmaps show LRD are UV-faint at fixed $L_{\text{H}\alpha}$ and dominate the most H$\alpha$-luminous objects at fixed $M_{\text{UV}}$ ($F356W$ for $z_{\text{spec}} \le 5$; $F444W$ for $z_{\text{spec}} > 5$).
Below, the Euler diagram and bar plot show that having a point source and a V-shape implies $\sim$80 percent odds of a broad line. 
Together, the RUBIES survey demonstrate that combining color-slope, compact-morphology, and spectroscopic-shape cuts can define LRD as a population \citep{hviding2025rubies}.
}
\label{fig:hviding}
\end{figure}

The characteristic bimodal spectral energy distribution was confirmed spectroscopically as one of the defining characteristics of LRD \citep{hviding2025rubies}. In Figure \ref{fig:hviding}, we present three complementary RUBIES diagnostics. First, the $\beta_{\rm UV}$–$\beta_{\rm opt}$ plane shows compact red sources in a tight locus distinct from the parent sample. Second, an Euler diagram quantifies the strong overlap among the three LRD criteria—broad Balmer lines, a dominant rest-optical point source, and a V-shaped continuum. Third, demographics linking redshift to JWST photometric filter flux ratio ratio F356W/F444W and UV absolute magnitude $M_{\rm UV}$ to $L_{\mathrm{H}\alpha}$ indicate that LRD are UV-faint at fixed $L_{\mathrm{H}\alpha}$ yet dominate among the most extreme ${\rm H}\alpha$ emitters. Taken together, these diagnostics tie the characteristic blue-UV/red-optical SED to an unresolved structure and broad Balmer emission, consistent with compact, partly obscured accretion.


\section{Observational Constraints on LRD}
\label{sec:obsConstraining}
Studying the rest-frame UV of LRD, detections of the high-ionization coronal line [Fe\small{X}] strongly suggest an AGN nature \citep{kocevski2023hidden,furtak2024high}. 
On the other hand, non-detections of He\small{II} emission lines in the far UV (FUV) and Mg\small{II} are more consistent with photoionization by massive stars \citep{akins2024strong}. 
Additionally, a Balmer break, observed for many LRD, typically suggests an aged stellar population, though it is not universal \citep{akins2024strong}. 
On the other hand, detections of strong C \textsc{iv}, marginal He \textsc{ii} and [Fe \textsc{x}], together with broad H$\alpha$ all combined strongly support an AGN interpretation \citep{inayoshi2025extremely}. 


As early as 2024, it was reasoned that (most) LRD dust masses must be quite limited $< 10^{4-5}\,M_\odot$ due to their compact sizes yet relatively low mass available for attenuation \citep{casey2024dust}. Indeed, deep NOEMA observations of $z > 7$ LRD \citep{xiao2025no} as well as multi-band Atacama Large Millimeter/Submillimeter Array \citep[ALMA;][]{wootten2009atacama} observations of a couple of the brightest LRD known resulted in stringent non-detections \citep{setton2025confirmed}. Observations of a larger sample ($\sim 60$) LRD with ALMA in the 1.3 mm continuum band similarly resulted in no detections, even with stacking, suggesting LRD either have modest dust reservoirs  ($A_v \sim 2-4$) or otherwise very dense gas $10^9\,\rm{cm}^{-3}$ causes the obscuration. The paucity of dust in LRD extends to a broader range of wavelengths in the far-IR and sub-mm: LRD observations in the rest-frame IR in Spitzer/MIPS 24 $\mu$m, JCMT/SCUBA-2 850 $\mu$m, ALMA 1.2 \& 2.0 mm, and MeerKAT/VLA 1.3 \& 3.0 GHz all resulted in non-detections \citep{bao2025investigating}.

Consistent with non-detections at redder wavelengths, \citet{brooks2025here} investigated the dust attenuation of Balmer narrow-line emission and broad-line signatures in 29 $z > 3.5$ AGN selected from JWST surveys (CEERS, JADES, and RUBIES). The narrow-line emission is consistent with no dust attenuation, which \citet{brooks2025here} consider to be consistent with star formation at larger scales in the galaxy. They place a lower limit on the dust attenuation $A_v > 3.6$ due to non-detection ($<3\sigma$) of the broad H$\beta$ line even after stacking 25 of their sources, which they say is consistent with a dust-obscured AGN.



Multi-wavelength observations resulting in non-detections have extended to an increasingly larger sample of LRD. Studying 434 LRD in the JWST COSMOS-Web survey, \citet{akins2025cosmos} find 
no detections, even when stacking, in X-ray, mid-IR, far-IR/submillimeter, and radio bands. 
Their non-detections are consistent with prior work of smaller samples that found LRD are not detected, or quite weakly emitting, in X-ray \citep{ananna2024x,yue2024stacking,maiolino2025jwst} and in radio \citep{mazzolari2024radio,perger2025deep,gloudemans2025another}.

\section{Theoretical Interpretations}
\label{sec:interp}
We next consider how LRD spectral properties arise and why AGN and stellar interpretations can both appear.
Forward modeling shows that an AGN continuum filtered by dense nearly dust free gas can generate a strong Balmer break and broad Balmer lines while leaving the UV relatively blue, a combination that matches several spectroscopic LRD with tentative variability present \citep[2.6$\sigma$;][]{d2025blackthunder}. 
In particular, the inclusion of mid-IR and redder data (from MIRI \& ALMA) has led to mixed results. 
While the red optical and near-IR data can be fit by an obscured accretion disk with scattered AGN light \citep{lambrides2024case,pacucci2024mildly,madau2025chasing}, stellar-dominated (even starburst) models provide improved fits \citep{perez2024nature,carranzaEscudero2025lonely}. 
Moreover, the concerns that a stellar nature of LRD is inconsistent with the maximal brightness surface density in the local universe \citep{hopkins2010maximum}, is alleviated due to MIRI and ALMA observations decreasing the requisite stellar mass \citep{williams2024galaxies}. 
On the other hand, recent analysis of LRD from the
JWST Advanced Deep Extragalactic Survey \citep[JADES;][]{eisenstein2023overview}
with NIRCam and MIRI photometry suggest one LRD with a 2\% AGN contribution to the luminosity, one completely star-forming, and the rest with 20-70\% contributions from AGN \citep{durodola2025exploring}. 
Similarly, recent analysis of MIRI data increases the fit for stellar components, reducing the number of AGN-only solutions without eliminating AGN altogether \citep{furtak2025investigating}. 

To explain the paucity of radio, X-ray, and far-IR/sub-mm emission, modeling work has extended beyond the standard models of AGN. 
Quasi star or black hole envelope models predict compact very red sources with strong Balmer features during specific evolutionary phases that overlap LRD selections, suggesting some LRD may represent heavy seeds in formation \citep{begelman2025little,durodola2025exploring}. 
Black hole envelopes embedded in dense gas can also reproduce red continua and strong lines in a way that mimics faint AGN \citep{kido2025black,naidu2025black}. 
On the other hand, simulations of collapsing clusters predict enhanced tidal disruption activity and luminous transients that can generate broad lines and variable continua in compact systems, offering another channel that suggests a non-standard AGN origin of LRD with rapid black hole growth in cluster cores \citep{bellovary2025little}. 
Dusty inflow models explain very red rest-optical colors and broad absorption and predict that mid-IR observations should be constraining \citep{li2025little}. 
Simulations of high redshift environments indicate that compact AGN can arise in gas rich protoclusters where repeated fueling and multiphase outflows sculpt line profiles and continuum shape
\citep{kannan2025introducing}. 
Observations of multiphase outflows around $z \sim 5$ quasars show that such winds can be common and energetic, which supports the idea that some LRD reflect wind-bearing accretion states rather than aged stellar populations \citep{brazzini2025multi}. These models collectively predict that modest changes in geometry, column density, and recent fueling history can toggle an object between apparently AGN-like and stellar-like diagnostics, which explains why LRD form a mixed class.

\begin{figure}[ht]
\begin{center}
\includegraphics[width=\textwidth]{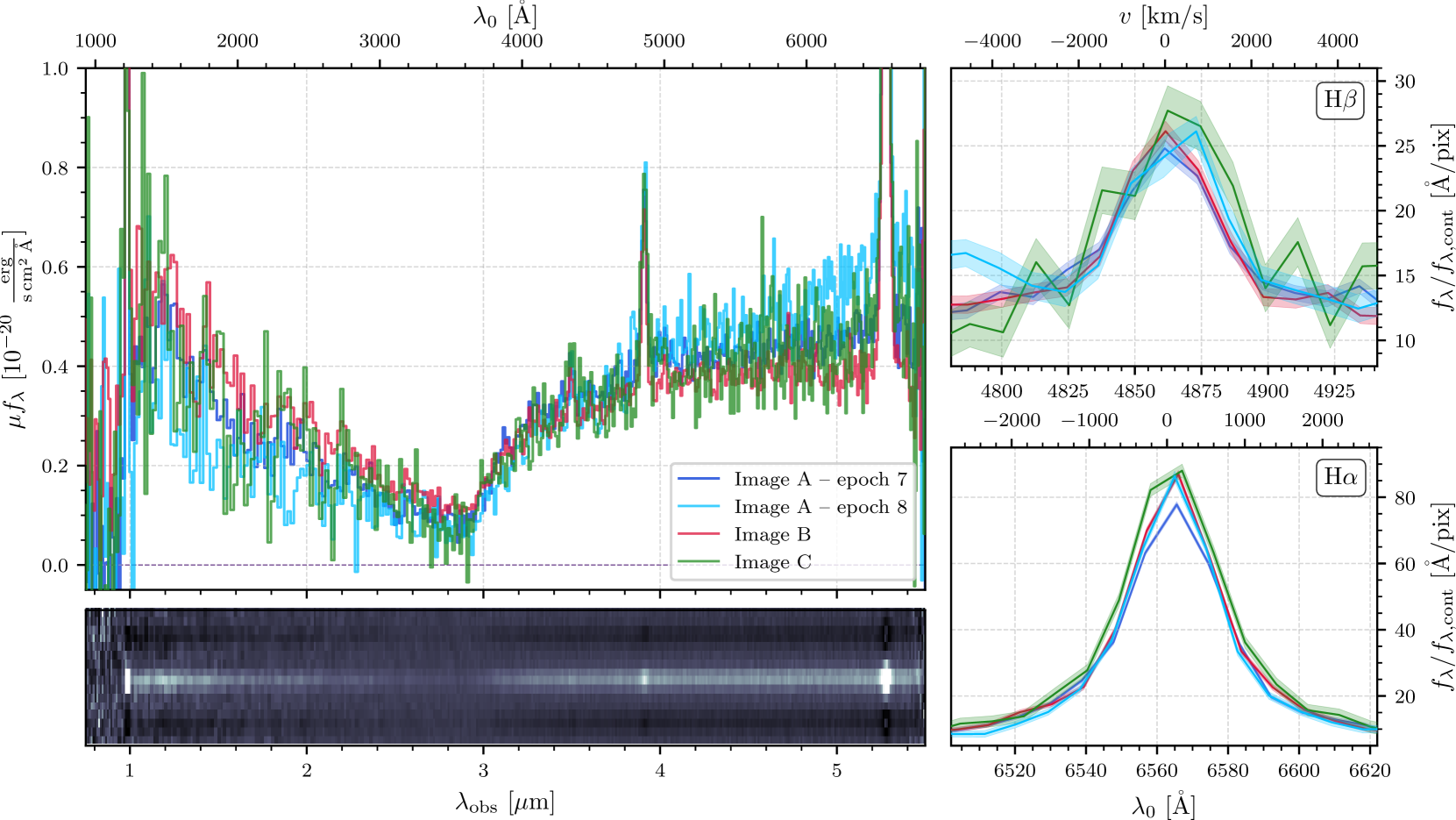}
\end{center}
\caption{
Figure credit: Lukas Furtak and Adi Zitrin, as adapted from \citet{furtak2025investigating}, their Figure 1. 
This figure provides direct spectroscopic evidence for accretion in the little red dot A2744-QSO1. JWST/NIRSpec-prism observations of the multiply imaged source A2744-QSO1 at multiple epochs capture the full Balmer region and reveal broad H$\alpha$ and H$\beta$ and their variability. After placing the spectra on a common reference and scaling line profiles by the local continua, the broad shapes persist while only modest equivalent-width changes remain, which is consistent with AGN variability. These measurements show that some LRD host broad-line regions even when continuum variability is weak, supporting an interpretation in which at least part of the LRD population traces active black hole growth \citep{furtak2025investigating}.}
\label{fig:furtak}
\end{figure}

Considering a multiply imaged LRD at redshift 7 from UNCOVER, \citet{furtak2025investigating} investigated time variability, leveraging lensing time delays of 22 years (rest-frame $\sim3\,$yr). Finding significant variability in H$\alpha$ and H$\beta$ lines, \citet{furtak2023jwst} confirmed the AGN nature of this LRD. We show \citet{furtak2025investigating} their figure 1 in Figure \ref{fig:furtak}. Clearly, the H$\alpha$ and H$\beta$ line fluxes have evolved significantly from epoch 7 (marked in blue) to epoch 8 (marked in green). This constitutes strong AGN evidence for several reasons. First, the Balmer lines are very broad (FWHM ~ a few $10^3kms^{-1}$), which requires the high velocities of gas near a black hole and cannot be produced by normal star-forming H \textsc{ii} regions. The line fluxes and profiles vary on rest-frame $\sim$year timescales, the expected behavior of a compact broad-line region responding to a changing ionizing source. Since the spectra are normalized and the multiple images are separated by known lensing delays, the differences cannot be due to calibration or lensing and must be intrinsic. Photometric non-variability does not contradict this result, since AGN continuum variability is stochastic and can be modest over the sampled interval. 

\section{Discussion and Conclusion}
\label{sec:disc_and_conclude}
Despite recent progress, many caveats surround the work performed to-date attempting to identify the nature of individual LRD. Many LRD are X-ray and radio quiet even in stacks, which could be consistent with heavy obscuration and low coronal efficiency (as expected from super-Eddington accretion models; \citealt{secunda2025little}) or purely stellar solutions; stellar templates fit a large fraction or LRD with mid-IR data included \citep{hainline2025investigation}. 
Radio and X-ray constraints specific to compact high redshift sources show that non detections do not rule out accretion for the luminosities expected from thick trapping dominated flows, which raises the need for sensitive follow up tuned to low efficiency spectra rather than relying on shallow stacks \citep{latif2025radio}. Deeper observations in radio bands may uncover a tail of detections if even a minority of LRD host jets or compact cores, which provides a near-term observational test of the mixed population picture \citep{latif2025synergies}. 

Completeness studies for obscured narrow line AGN show that color cuts and emission line thresholds can miss faint or partially obscured accretors, which implies that some LRD-like systems may be absent from current AGN catalogs and that selection bias can skew interpretations \citep{bouwens2025beneath,scholtz2025jades}. 

Ultimately, LRD likely comprise both genuine rapidly growing black holes and compact star-forming systems, and resolving their roles in early black hole demographics will require deeper mid-IR coverage, higher signal to noise spectroscopy that targets both Balmer and high ionization UV lines, and radio and X-ray follow up that is matched to the efficiencies expected for thick discs rather than thin ones \citep{akins2025tentative}.

In particular, measuring time variability would provide `smoking-gun' evidence for an AGN nature of individual LRD. As most LRD are not multiply imaged as in the target of \citet{furtak2025investigating}, collecting additional epochs of LRD photometric and spectroscopic measurements extending several years will be necessary. The upcoming \texttt{TWINKLE} campaign (JWST Proposal Cycle 4, ID 7404PI, PI: Naidu) will be the next possibility to find variability in the Balmer emission line fluxes.
While expensive, deeper observations of truly AGN-generating LRD will be crucial to better understand the assembly of supermassive black holes within the first billion years from the Big Bang.



\section*{Conflict of Interest Statement}

The authors declare that the research was conducted in the absence of any commercial or financial relationships that could be construed as a potential conflict of interest.

\section*{Author Contributions}
DDV: Writing -- original draft, review and editing. RJF: Writing -- review and editing.

\section*{Acknowledgments}
We thank Lukas Furtak and Adi Zitrin for their permission to re-use their figure as shown in our Figure \ref{fig:furtak}.
We thank Raphael Hviding for their permission to re-use their figure as shown in our Figure \ref{fig:hviding}.
DDV thanks his family for their belief and support throughout the writing process.

\bibliographystyle{Frontiers-Harvard} 
\bibliography{test}

@article{graham2025dot,
  title={Dot to dot: High-z little red dots in diagrams with galaxy-morphology-specific scaling relations},
  author={Graham, Alister W and Chilingarian, Igor and Nguyen, Dieu Duc and Soria, Roberto and Durr{\'e}, Mark and Forbes, Duncan A},
  journal={Publications of the Astronomical Society of Australia},
  volume={42},
  pages={e068},
  year={2025},
  publisher={Cambridge University Press}
}

@article{d2025blackthunder,
  title={BlackTHUNDER strikes twice: rest-frame Balmer-line absorption and high Eddington accretion rate in a Little Red Dot at $ z= 7.04$},
  author={D'Eugenio, Francesco and Maiolino, Roberto and Perna, Michele and Uebler, Hannah and Ji, Xihan and McClymont, William and Koudmani, Sophie and Sijacki, Debora and Juod{\v{z}}balis, Ignas and Scholtz, Jan and others},
  journal={arXiv preprint arXiv:2503.11752},
  year={2025}
}

@article{durodola2025exploring,
  title={Exploring the Active Galactic Nucleus Fraction of a Sample of JWST’s Little Red Dots at 4< z< 8: Overmassive Black Holes Are Strongly Favored},
  author={Durodola, Emmanuel and Pacucci, Fabio and Hickox, Ryan C},
  journal={The Astrophysical Journal},
  volume={985},
  number={2},
  pages={169},
  year={2025},
  publisher={IOP Publishing}
}

@article{mazzolari2024radio,
  title={The radio properties of the JWST-discovered AGN},
  author={Mazzolari, G and Gilli, R and Maiolino, R and Prandoni, I and Delvecchio, I and Norman, C and Jimenez-Andrade, EF and Belladitta, S and Vito, F and Momjian, E and others},
  journal={arXiv preprint arXiv:2412.04224},
  year={2024}
}

@article{maiolino2025jwst,
  title={JWST meets Chandra: a large population of Compton thick, feedback-free, and intrinsically X-ray weak AGN, with a sprinkle of SNe},
  author={Maiolino, Roberto and Risaliti, Guido and Signorini, Matilde and Trefoloni, Bartolomeo and Juod{\v{z}}balis, Ignas and Scholtz, Jan and {\"U}bler, Hannah and D’Eugenio, Francesco and Carniani, Stefano and Fabian, Andy and others},
  journal={Monthly Notices of the Royal Astronomical Society},
  volume={538},
  number={3},
  pages={1921--1943},
  year={2025},
  publisher={Oxford University Press}
}

@article{eisenstein2023overview,
  title={Overview of the JWST advanced deep extragalactic survey (JADES)},
  author={Eisenstein, Daniel J and Willott, Chris and Alberts, Stacey and Arribas, Santiago and Bonaventura, Nina and Bunker, Andrew J and Cameron, Alex J and Carniani, Stefano and Charlot, Stephane and Curtis-Lake, Emma and others},
  journal={arXiv preprint arXiv:2306.02465},
  year={2023}
}

@article{secunda2025little,
  title={Do Little Red Dots Vary?},
  author={Secunda, Amy and Somerville, Rachel S and Jiang, Yan-Fei and Greene, Jenny E and Furtak, Lukas J and Zitrin, Adi},
  journal={The Astrophysical Journal},
  volume={996},
  number={1},
  pages={6},
  year={2025},
  publisher={IOP Publishing}
}

@article{furtak2023jwst,
  title={JWST UNCOVER: extremely red and compact object at z phot 7.6 triply imaged by A2744},
  author={Furtak, Lukas J and Zitrin, Adi and Plat, Ad{\`e}le and Fujimoto, Seiji and Wang, Bingjie and Nelson, Erica J and Labb{\'e}, Ivo and Bezanson, Rachel and Brammer, Gabriel B and van Dokkum, Pieter and others},
  journal={The Astrophysical Journal},
  volume={952},
  number={2},
  pages={142},
  year={2023},
  publisher={IOP Publishing}
}

@article{furtak2024high,
  title={A high black-hole-to-host mass ratio in a lensed AGN in the early Universe},
  author={Furtak, Lukas J and Labb{\'e}, Ivo and Zitrin, Adi and Greene, Jenny E and Dayal, Pratika and Chemerynska, Iryna and Kokorev, Vasily and Miller, Tim B and Goulding, Andy D and de Graaff, Anna and others},
  journal={Nature},
  volume={628},
  number={8006},
  pages={57--61},
  year={2024},
  publisher={Nature Publishing Group UK London}
}

@article{furtak2025investigating,
  title={Investigating photometric and spectroscopic variability in the multiply imaged little red dot A2744-QSO1},
  author={Furtak, Lukas J and Secunda, Amy R and Greene, Jenny E and Zitrin, Adi and Labb{\'e}, Ivo and Golubchik, Miriam and Bezanson, Rachel and Kokorev, Vasily and Atek, Hakim and Brammer, Gabriel B and others},
  journal={Astronomy \& Astrophysics},
  volume={698},
  pages={A227},
  year={2025},
  publisher={EDP Sciences}
}

@article{casey2024dust,
  title={Dust in Little Red Dots},
  author={Casey, Caitlin M and Akins, Hollis B and Kokorev, Vasily and McKinney, Jed and Cooper, Olivia R and Long, Arianna S and Franco, Maximilien and Manning, Sinclaire M},
  journal={The Astrophysical Journal Letters},
  volume={975},
  number={1},
  pages={L4},
  year={2024},
  publisher={IOP Publishing}
}

@article{perger2025deep,
  title={Deep silence: Radio properties of little red dots},
  author={Perger, K and Fogasy, J and Frey, S and Gab{\'a}nyi, K{\'E}},
  journal={Astronomy \& Astrophysics},
  volume={693},
  pages={L2},
  year={2025},
  publisher={EDP Sciences}
}

@article{gloudemans2025another,
  title={Another Piece to the Puzzle: Radio Detection of a JWST-detected Active Galactic Nucleus Candidate},
  author={Gloudemans, Anniek J and Duncan, Kenneth J and Eilers, Anna-Christina and Farina, Emanuele Paolo and Harikane, Yuichi and Inayoshi, Kohei and Lambrides, Erini and Vardoulaki, Eleni},
  journal={The Astrophysical Journal},
  volume={986},
  number={2},
  pages={130},
  year={2025},
  publisher={IOP Publishing}
}

@article{setton2025confirmed,
  title={A confirmed deficit of hot and cold dust emission in the most luminous Little Red Dots},
  author={Setton, David J and Greene, Jenny E and Spilker, Justin S and Williams, Christina C and Labbe, Ivo and Ma, Yilun and Wang, Bingjie and Whitaker, Katherine E and Leja, Joel and de Graaff, Anna and others},
  journal={arXiv preprint arXiv:2503.02059},
  year={2025}
}

@article{akins2025cosmos,
  title={COSMOS-web: The Overabundance and Physical Nature of “Little Red Dots”—Implications for Early Galaxy and SMBH Assembly},
  author={Akins, Hollis B and Casey, Caitlin M and Lambrides, Erini and Allen, Natalie and Andika, Irham T and Brinch, Malte and Champagne, Jaclyn B and Cooper, Olivia and Ding, Xuheng and Drakos, Nicole E and others},
  journal={The Astrophysical Journal},
  volume={991},
  number={1},
  pages={37},
  year={2025},
  publisher={IOP Publishing}
}

@article{akins2025tentative,
  title={Tentative detection of neutral gas in a Little Red Dot at $ z= 4.46$},
  author={Akins, Hollis B and Casey, Caitlin M and Chisholm, John and Berg, Danielle A and Cooper, Olivia and Franco, Maximilien and Fujimoto, Seiji and Lambrides, Erini and Long, Arianna S and McKinney, Jed},
  journal={arXiv preprint arXiv:2503.00998},
  year={2025}
}

@article{hopkins2010maximum,
  title={A maximum stellar surface density in dense stellar systems},
  author={Hopkins, Philip F and Murray, Norman and Quataert, Eliot and Thompson, Todd A},
  journal={Monthly Notices of the Royal Astronomical Society: Letters},
  volume={401},
  number={1},
  pages={L19--L23},
  year={2010},
  publisher={The Royal Astronomical Society}
}

@article{labbe2023population,
  title={A population of red candidate massive galaxies\~{} 600 Myr after the Big Bang},
  author={Labb{\'e}, Ivo and van Dokkum, Pieter and Nelson, Erica and Bezanson, Rachel and Suess, Katherine A and Leja, Joel and Brammer, Gabriel and Whitaker, Katherine and Mathews, Elijah and Stefanon, Mauro and others},
  journal={Nature},
  volume={616},
  number={7956},
  pages={266--269},
  year={2023},
  publisher={Nature Publishing Group UK London}
}

@article{barro2024extremely,
  title={Extremely red galaxies at z= 5--9 with MIRI and NIRSpec: dusty galaxies or obscured active galactic nuclei?},
  author={Barro, Guillermo and P{\'e}rez-Gonz{\'a}lez, Pablo G and Kocevski, Dale D and McGrath, Elizabeth J and Trump, Jonathan R and Simons, Raymond C and Somerville, Rachel S and Yung, LY Aaron and Haro, Pablo Arrabal and Akins, Hollis B and others},
  journal={The Astrophysical Journal},
  volume={963},
  number={2},
  pages={128},
  year={2024},
  publisher={IOP Publishing}
}

@article{labbe2024uncover,
  title={UNCOVER: candidate red active galactic nuclei at 3< z< 7 with JWST and ALMA},
  author={Labbe, Ivo and Greene, Jenny E and Bezanson, Rachel and Fujimoto, Seiji and Furtak, Lukas J and Goulding, Andy D and Matthee, Jorryt and Naidu, Rohan P and Oesch, Pascal A and Atek, Hakim and others},
  journal={The Astrophysical Journal},
  volume={978},
  number={1},
  pages={92},
  year={2024},
  publisher={IoP Publishing}
}

@article{hainline2025investigation,
  title={An investigation into the selection and colors of little red dots and active galactic nuclei},
  author={Hainline, Kevin N and Maiolino, Roberto and Juod{\v{z}}balis, Ignas and Scholtz, Jan and {\"U}bler, Hannah and d’Eugenio, Francesco and Helton, Jakob M and Sun, Yang and Sun, Fengwu and Robertson, Brant and others},
  journal={The Astrophysical Journal},
  volume={979},
  number={2},
  pages={138},
  year={2025},
  publisher={IOP Publishing}
}

@article{bezanson2024jwst,
  title={The JWST UNCOVER Treasury survey: ultradeep NIRSpec and NIRCam observations before the epoch of reionization},
  author={Bezanson, Rachel and Labbe, Ivo and Whitaker, Katherine E and Leja, Joel and Price, Sedona H and Franx, Marijn and Brammer, Gabriel and Marchesini, Danilo and Zitrin, Adi and Wang, Bingjie and others},
  journal={The Astrophysical Journal},
  volume={974},
  number={1},
  pages={92},
  year={2024},
  publisher={IOP Publishing}
}

@article{kashino2023eiger,
  title={EIGER. I. A large sample of [O iii]-emitting galaxies at 5.3< z< 6.9 and direct evidence for local reionization by galaxies},
  author={Kashino, Daichi and Lilly, Simon J and Matthee, Jorryt and Eilers, Anna-Christina and Mackenzie, Ruari and Bordoloi, Rongmon and Simcoe, Robert A},
  journal={The Astrophysical Journal},
  volume={950},
  number={1},
  pages={66},
  year={2023},
  publisher={IOP Publishing}
}

@article{oesch2023jwst,
  title={The JWST FRESCO survey: legacy NIRCam/grism spectroscopy and imaging in the two GOODS fields},
  author={Oesch, Pascal A and Brammer, G and Naidu, RP and Bouwens, RJ and Chisholm, John and Illingworth, GD and Matthee, Jorryt and Nelson, E and Qin, Y and Reddy, N and others},
  journal={Monthly Notices of the Royal Astronomical Society},
  volume={525},
  number={2},
  pages={2864--2874},
  year={2023},
  publisher={Oxford University Press}
}

@article{mcelwain2023james,
  title={The James Webb Space Telescope Mission: optical telescope element design, development, and performance},
  author={McElwain, Michael W and Feinberg, Lee D and Perrin, Marshall D and Clampin, Mark and Mountain, C Matt and Lallo, Matthew D and Lajoie, Charles-Philippe and Kimble, Randy A and Bowers, Charles W and Stark, Christopher C and others},
  journal={Publications of the Astronomical Society of the Pacific},
  volume={135},
  number={1047},
  pages={058001},
  year={2023},
  publisher={IOP Publishing}
}

@article{gardner2023james,
       author = {{Gardner}, Jonathan P. and {Mather}, John C. and {Abbott}, Randy and {Abell}, James S. and {Abernathy}, Mark and {Abney}, Faith E. and {Abraham}, John G. and {Abraham}, Roberto and {Abul-Huda}, Yasin M. and {Acton}, Scott and {Adams}, Cynthia K. and {Adams}, Evan and {Adler}, David S. and {Adriaensen}, Maarten and {Aguilar}, Jonathan Albert and {Ahmed}, Mansoor and {Ahmed}, Nasif S. and {Ahmed}, Tanjira and {Albat}, R{\"u}deger and {Albert}, Lo{\"\i}c and {Alberts}, Stacey and {Aldridge}, David and {Allen}, Mary Marsha and {Allen}, Shaune S. and {Altenburg}, Martin and {Altunc}, Serhat and {Alvarez}, Jose Lorenzo and {{\'A}lvarez-M{\'a}rquez}, Javier and {Alves de Oliveira}, Catarina and {Ambrose}, Leslie L. and {Anandakrishnan}, Satya M. and {Andersen}, Gregory C. and {Anderson}, Harry James and {Anderson}, Jay and {Anderson}, Kristen and {Anderson}, Sara M. and {Aprea}, Julio and {Archer}, Benita J. and {Arenberg}, Jonathan W. and {Argyriou}, Ioannis and {Arribas}, Santiago and {Artigau}, {\'E}tienne and {Arvai}, Amanda Rose and {Atcheson}, Paul and {Atkinson}, Charles B. and {Averbukh}, Jesse and {Aymergen}, Cagatay and {Bacinski}, John J. and {Baggett}, Wayne E. and {Bagnasco}, Giorgio and {Baker}, Lynn L. and {Balzano}, Vicki Ann and {Banks}, Kimberly A. and {Baran}, David A. and {Barker}, Elizabeth A. and {Barrett}, Larry K. and {Barringer}, Bruce O. and {Barto}, Allison and {Bast}, William and {Baudoz}, Pierre and {Baum}, Stefi and {Beatty}, Thomas G. and {Beaulieu}, Mathilde and {Bechtold}, Kathryn and {Beck}, Tracy and {Beddard}, Megan M. and {Beichman}, Charles and {Bellagama}, Larry and {Bely}, Pierre and {Berger}, Timothy W. and {Bergeron}, Louis E. and {Bernier}, Antoine-Darveau and {Bertch}, Maria D. and {Beskow}, Charlotte and {Betz}, Laura E. and {Biagetti}, Carl P. and {Birkmann}, Stephan and {Bjorklund}, Kurt F. and {Blackwood}, James D. and {Blazek}, Ronald Paul and {Blossfeld}, Stephen and {Bluth}, Marcel and {Boccaletti}, Anthony and {Boegner}, Jr., Martin E. and {Bohlin}, Ralph C. and {Boia}, John Joseph and {B{\"o}ker}, Torsten and {Bonaventura}, N. and {Bond}, Nicholas A. and {Bosley}, Kari Ann and {Boucarut}, Rene A. and {Bouchet}, Patrice and {Bouwman}, Jeroen and {Bower}, Gary and {Bowers}, Ariel S. and {Bowers}, Charles W. and {Boyce}, Leslye A. and {Boyer}, Christine T. and {Boyer}, Martha L. and {Boyer}, Michael and {Boyer}, Robert and {Bradley}, Larry D. and {Brady}, Gregory R. and {Brandl}, Bernhard R. and {Brannen}, Judith L. and {Breda}, David and {Bremmer}, Harold G. and {Brennan}, David and {Bresnahan}, Pamela A. and {Bright}, Stacey N. and {Broiles}, Brian J. and {Bromenschenkel}, Asa and {Brooks}, Brian H. and {Brooks}, Keira J. and {Brown}, Bob and {Brown}, Bruce and {Brown}, Thomas M. and {Bruce}, Barry W. and {Bryson}, Jonathan G. and {Bujanda}, Edwin D. and {Bullock}, Blake M. and {Bunker}, A.~J. and {Bureo}, Rafael and {Burt}, Irving J. and {Bush}, James Aaron and {Bushouse}, Howard A. and {Bussman}, Marie C. and {Cabaud}, Olivier and {Cale}, Steven and {Calhoon}, Charles D. and {Calvani}, Humberto and {Canipe}, Alicia M. and {Caputo}, Francis M. and {Cara}, Mihai and {Carey}, Larkin and {Case}, Michael Eli and {Cesari}, Thaddeus and {Cetorelli}, Lee D. and {Chance}, Don R. and {Chandler}, Lynn and {Chaney}, Dave and {Chapman}, George N. and {Charlot}, S. and {Chayer}, Pierre and {Cheezum}, Jeffrey I. and {Chen}, Bin and {Chen}, Christine H. and {Cherinka}, Brian and {Chichester}, Sarah C. and {Chilton}, Zachary S. and {Chittiraibalan}, Dharini and {Clampin}, Mark and {Clark}, Charles R. and {Clark}, Kerry W. and {Clark}, Stephanie M. and {Claybrooks}, Edward E. and {Cleveland}, Keith A. and {Cohen}, Andrew L. and {Cohen}, Lester M. and {Col{\'o}n}, Knicole D. and {Coleman}, Benee L. and {Colina}, Luis and {Comber}, Brian J. and {Comeau}, Thomas M. and {Comer}, Thomas and {Conde Reis}, Alain and {Connolly}, Dennis C. and {Conroy}, Kyle E. and {Contos}, Adam R. and {Contreras}, James and {Cook}, Neil J. and {Cooper}, James L. and {Cooper}, Rachel Aviva and {Correia}, Michael F. and {Correnti}, Matteo and {Cossou}, Christophe and {Costanza}, Brian F. and {Coulais}, Alain and {Cox}, Colin R. and {Coyle}, Ray T. and {Cracraft}, Misty M. and {Crew}, Keith A. and {Curtis}, Gary J. and {Cusveller}, Bianca and {Da Costa Maciel}, Cleyciane and {Dailey}, Christopher T. and {Daugeron}, Fr{\'e}d{\'e}ric and {Davidson}, Greg S. and {Davies}, James E. and {Davis}, Katherine Anne and {Davis}, Michael S. and {Day}, Ratna and {de Chambure}, Daniel and {de Jong}, Pauline and {De Marchi}, Guido and {Dean}, Bruce H. and {Decker}, John E. and {Delisa}, Amy S. and {Dell}, Lawrence C. and {Dellagatta}, Gail},
        title = "{The James Webb Space Telescope Mission}",
      journal = {Publications of the Astronomical Society of the Pacific},
         year = 2023,
        month = jun,
       volume = {135},
       number = {1048},
          eid = {068001},
        pages = {068001},
          doi = {10.1088/1538-3873/acd1b5},
archivePrefix = {arXiv},
       eprint = {2304.04869},
 primaryClass = {astro-ph.IM},
       adsurl = {https://ui.adsabs.harvard.edu/abs/2023PASP..135f8001G}
}

@article{naidu2025black,
  title={A" Black Hole Star" Reveals the Remarkable Gas-Enshrouded Hearts of the Little Red Dots},
  author={Naidu, Rohan P and Matthee, Jorryt and Katz, Harley and de Graaff, Anna and Oesch, Pascal and Smith, Aaron and Greene, Jenny E and Brammer, Gabriel and Weibel, Andrea and Hviding, Raphael and others},
  journal={arXiv preprint arXiv:2503.16596},
  year={2025}
}

@article{kido2025black,
  title={Black Hole Envelopes in Little Red Dots},
  author={Kido, Daisaburo and Ioka, Kunihito and Hotokezaka, Kenta and Inayoshi, Kohei and Irwin, Christopher M},
  journal={arXiv preprint arXiv:2505.06965},
  year={2025}
}

@article{bellovary2025little,
  title={Little Red Dots Are Tidal Disruption Events in Runaway-collapsing Clusters},
  author={Bellovary, Jillian},
  journal={The Astrophysical Journal Letters},
  volume={984},
  number={2},
  pages={L55},
  year={2025},
  publisher={IOP Publishing}
}

@article{kannan2025introducing,
  title={Introducing the THESAN-ZOOM project: radiation-hydrodynamic simulations of high-redshift galaxies with a multi-phase interstellar medium},
  author={Kannan, Rahul and Puchwein, Ewald and Smith, Aaron and Borrow, Josh and Garaldi, Enrico and Keating, Laura and Vogelsberger, Mark and Zier, Oliver and McClymont, William and Shen, Xuejian and others},
  journal={arXiv preprint arXiv:2502.20437},
  year={2025}
}

@article{scholtz2025jades,
  title={JADES: A large population of obscured, narrow-line active galactic nuclei at high redshift},
  author={Scholtz, Jan and Maiolino, Roberto and D’Eugenio, Francesco and Curtis-Lake, Emma and Carniani, Stefano and Charlot, Stephane and Curti, Mirko and Silcock, Maddie S and Arribas, Santiago and Baker, William and others},
  journal={Astronomy \& Astrophysics},
  volume={697},
  pages={A175},
  year={2025},
  publisher={EDP Sciences}
}

@article{bouwens2025beneath,
  title={Beneath the Surface:> 85\% of z> 5.9 QSOs in Massive Host Galaxies are UV-Faint},
  author={Bouwens, Rychard J and Banados, Eduardo and Decarli, Roberto and Hennawi, Joseph and Yang, Daming and Algera, Hiddo and Aravena, Manuel and Farina, Emanuele and Gloudemans, Anniek and Hodge, Jacqueline and others},
  journal={arXiv preprint arXiv:2506.24128},
  year={2025}
}

@article{hviding2025rubies,
  title={RUBIES: A Spectroscopic Census of Little Red Dots; All V-Shaped Point Sources Have Broad Lines},
  author={Hviding, Raphael E and de Graaff, Anna and Miller, Tim B and Setton, David J and Greene, Jenny E and Labb{\'e}, Ivo and Brammer, Gabriel and Bezanson, Rachel and Boogaard, Leindert A and Cleri, Nikko J and others},
  journal={arXiv preprint arXiv:2506.05459},
  year={2025}
}

@article{brooks2025here,
  title={Here There Be (Dusty) Monsters: High-redshift Active Galactic Nuclei Are Dustier than Their Hosts},
  author={Brooks, Madisyn and Simons, Raymond C and Trump, Jonathan R and Taylor, Anthony J and Bagley, Micaela B and Backhaus, Bren and Davis, Kelcey and Buat, V{\'e}ronique and Cleri, Nikko J and de la Vega, Alexander and others},
  journal={The Astrophysical Journal},
  volume={986},
  number={2},
  pages={177},
  year={2025},
  publisher={IOP Publishing}
}

@article{kocevski2023hidden,
  title={Hidden little monsters: spectroscopic identification of low-mass, broad-line AGNs at z> 5 with CEERS},
  author={Kocevski, Dale D and Onoue, Masafusa and Inayoshi, Kohei and Trump, Jonathan R and Haro, Pablo Arrabal and Grazian, Andrea and Dickinson, Mark and Finkelstein, Steven L and Kartaltepe, Jeyhan S and Hirschmann, Michaela and others},
  journal={The Astrophysical Journal Letters},
  volume={954},
  number={1},
  pages={L4},
  year={2023},
  publisher={IOP Publishing}
}

@article{wootten2009atacama,
  title={The Atacama large millimeter/submillimeter array},
  author={Wootten, Alwyn and Thompson, A Richard},
  journal={Proceedings of the IEEE},
  volume={97},
  number={8},
  pages={1463--1471},
  year={2009},
  publisher={IEEE}
}

@article{onoue2023candidate,
  title={A Candidate for the least-massive black hole in the first 1.1 billion years of the universe},
  author={Onoue, Masafusa and Inayoshi, Kohei and Ding, Xuheng and Li, Wenxiu and Li, Zhengrong and Molina, Juan and Inoue, Akio K and Jiang, Linhua and Ho, Luis C},
  journal={The Astrophysical Journal Letters},
  volume={942},
  number={1},
  pages={L17},
  year={2023},
  publisher={IOP Publishing}
}

@article{begelman2025little,
  title={Little Red Dots As Late-stage Quasi-stars},
  author={Begelman, Mitchell C and Dexter, Jason},
  journal={arXiv preprint arXiv:2507.09085},
  year={2025}
}

@article{carnall2023massive,
  title={A massive quiescent galaxy at redshift 4.658},
  author={Carnall, Adam C and McLure, Ross J and Dunlop, James S and McLeod, Derek J and Wild, Vivienne and Cullen, Fergus and Magee, Dan and Begley, Ryan and Cimatti, Andrea and Donnan, Callum T and others},
  journal={Nature},
  volume={619},
  number={7971},
  pages={716--719},
  year={2023},
  publisher={Nature Publishing Group UK London}
}

@article{rieke2023performance,
  title={Performance of NIRCam on JWST in flight},
  author={Rieke, Marcia J and Kelly, Douglas M and Misselt, Karl and Stansberry, John and Boyer, Martha and Beatty, Thomas and Egami, Eiichi and Florian, Michael and Greene, Thomas P and Hainline, Kevin and others},
  journal={Publications of the Astronomical Society of the Pacific},
  volume={135},
  number={1044},
  pages={028001},
  year={2023},
  publisher={IOP Publishing}
}

@article{argyriou2023jwst,
  title={JWST MIRI flight performance: the medium-resolution spectrometer},
  author={Argyriou, Ioannis and Glasse, Alistair and Law, David R and Labiano, Alvaro and {\'A}lvarez-M{\'a}rquez, Javier and Patapis, Polychronis and Kavanagh, Patrick J and Gasman, Danny and Mueller, Michael and Larson, Kirsten and others},
  journal={Astronomy \& Astrophysics},
  volume={675},
  pages={A111},
  year={2023},
  publisher={EDP Sciences}
}

@article{finkelstein2023ceers,
  title={CEERS key paper. I. An early look into the first 500 Myr of galaxy formation with JWST},
  author={Finkelstein, Steven L and Bagley, Micaela B and Ferguson, Henry C and Wilkins, Stephen M and Kartaltepe, Jeyhan S and Papovich, Casey and Yung, LY Aaron and Haro, Pablo Arrabal and Behroozi, Peter and Dickinson, Mark and others},
  journal={The Astrophysical journal letters},
  volume={946},
  number={1},
  pages={L13},
  year={2023},
  publisher={IOP Publishing}
}

@article{akins2024strong,
  title={Strong rest-UV emission lines in a" little red dot" AGN at $ z= 7$: Early SMBH growth alongside compact massive star formation?},
  author={Akins, Hollis B and Casey, Caitlin M and Berg, Danielle A and Chisholm, John and Franco, Maximilien and Finkelstein, Steven L and Fujimoto, Seiji and Kokorev, Vasily and Lambrides, Erini and Robertson, Brant E and others},
  journal={arXiv preprint arXiv:2410.00949},
  year={2024}
}

@article{yue2024stacking,
  title={Stacking X-ray observations of “Little Red Dots”: Implications for their active galactic nucleus properties},
  author={Yue, Minghao and Eilers, Anna-Christina and Ananna, Tonima Tasnim and Panagiotou, Christos and Kara, Erin and Miyaji, Takamitsu},
  journal={The Astrophysical Journal Letters},
  volume={974},
  number={2},
  pages={L26},
  year={2024},
  publisher={IOP Publishing}
}


\section*{Figure captions}


\end{document}